\pdfoutput=1
\RequirePackage{ifpdf}
\ifpdf 
\documentclass[pdftex]{sigma}
\else
\documentclass{sigma}
\fi

\numberwithin{equation}{section}

\begin{document}

\allowdisplaybreaks

\newcommand{\arXivNumber}{1703.07731}

\renewcommand{\thefootnote}{}

\renewcommand{\PaperNumber}{056}

\FirstPageHeading

\ShortArticleName{Zero Range Process and Multi-Dimensional Random Walks}

\ArticleName{Zero Range Process and Multi-Dimensional\\ Random Walks\footnote{This paper is a~contribution to the Special Issue on Recent Advances in Quantum Integrable Systems. The full collection is available at \href{http://www.emis.de/journals/SIGMA/RAQIS2016.html}{http://www.emis.de/journals/SIGMA/RAQIS2016.html}}}

\Author{Nicolay M.~BOGOLIUBOV~$^{\dag\ddag}$ and Cyril MALYSHEV~$^{\dag\ddag}$}

\AuthorNameForHeading{N.M.~Bogoliubov and C.~Malyshev}

\Address{$^\dag$~St.-Petersburg Department of Steklov Institute of Mathematics of RAS,\\
\hphantom{$^\dag$}~Fontanka 27, St.-Petersburg, Russia}
\EmailD{\href{mailto:bogoliub@yahoo.com}{bogoliub@yahoo.com}, \href{mailto:malyshev@pdmi.ras.ru}{malyshev@pdmi.ras.ru}}

\Address{$^{\ddag}$ ITMO University, Kronverksky 49, St.-Petersburg, Russia}

\ArticleDates{Received March 28, 2017, in f\/inal form July 14, 2017; Published online July 22, 2017}

\vspace{-1mm}

\Abstract{The special limit of the totally asymmetric zero range process of the low-dimen\-sional non-equilibrium statistical mechanics described by the non-Hermitian Hamiltonian is considered. The calculation of the conditional probabilities of the model are based on the algebraic Bethe ansatz approach. We demonstrate that the conditional probabilities may be considered as the generating functions of the random multi-dimensional lattice walks bounded by a hyperplane. This type of walks we call the walks over the multi-dimensional simplicial lattices. The answers for the conditional probability and for the number of random walks in the multi-dimensional simplicial lattice are expressed through the symmetric functions.}


\Classification{05A19; 05E05; 82B23}

\renewcommand{\thefootnote}{\arabic{footnote}}
\setcounter{footnote}{0}

\vspace{-2.5mm}

\section{Introduction}

The zero-range process is a stochastic lattice gas where the particles hop randomly with an on-site interaction that makes the jump rate dependent only on the local particle number. The zero range processes (ZRPs) belong to a class of minimal statistical-mechanics models of the low-dimensional non-equilibrium physics \cite{evans, spit}. Being exactly solvable the model and its other variations are intensively studied both by mathematicians and physicists \cite{brak,spohn,kanai,kuniba,povolo,povol}.

In this paper we consider the totally asymmetric simple zero range process (TASZRP) \cite{f,kbi}, which describes a system of indistinguishable particles placed on a one-dimensional lattice, moving randomly in one direction from right to left with the equal hopping rate on a~periodic ring. The dynamical variables of the model are the \textit{phase operators} \cite{can} which can be regarded as a special limit of $q$-bosons \cite{bbp, kd}. The application of the quantum inverse method (QIM) \cite{f,kbi,kul} allows to calculate the scalar products and form-factors of the model and represent them in the determinantal form \cite{bogtmf, nas}. The relation of the considered model and the totaly asymmetric simple exclusion process (TASEP) was discussed in \cite{nas,mendes}.

Certain quantum integrable models solvable by the QIM demonstrate close relationship \cite{b, xxz3, bmumn} with the dif\/ferent objects of the enumerative combinatorics \cite{stan1,stan2} and the theory of the symmetric functions \cite{macd}. It appeared that the correlation functions of some integrable models may be regarded as the generating functions of plane partitions and random walks.

Dif\/ferent types of random walks \cite{fors,kratt, stan1,stan2} are of considerable recent interest due to their role in quantum information processing \cite{shor,fred}. The walks on multi-dimensional lattices were studied by many authors \cite{bogspr, mdm,mortimer,mdr}.

In this paper we shall calculate the conditional probability of the model and reveal its connection with the generating function of the lattice paths
on multi-dimensional oriented lattices bounded by a hyperplane.

The layout of the paper is as follows. In the introductory Section~\ref{section2} we give the def\/inition of the TASZRP. The solution of the model by QIM is presented in Section~\ref{section3}. In Section~\ref{section4} conditional probability is calculated. The random walks over the $M$-dimensional oriented simplicial lattice with retaining boundary conditions are introduced and the connection of their generating function with the conditional probability is established.

\section{Totally asymmetric simple zero range hopping model}\label{section2}

Consider a system with $N$ particles on a periodic one-dimensional lattice of length $M+1$, i.e., on a ring, where sites $m$ and $m+M+1$ are identical. Each site of a~lattice contains arbitrary number of particles. The particles evolve with the following dynamics rule: during the time interval [$t,t+{\rm d}t$] a~particle on a site $i$ jumps with probability $dt$ to the neighbouring site $i+1$. There are no restrictions on the number of particles on a lattice site.
\begin{figure}[h]\centering
\includegraphics [scale=.3]{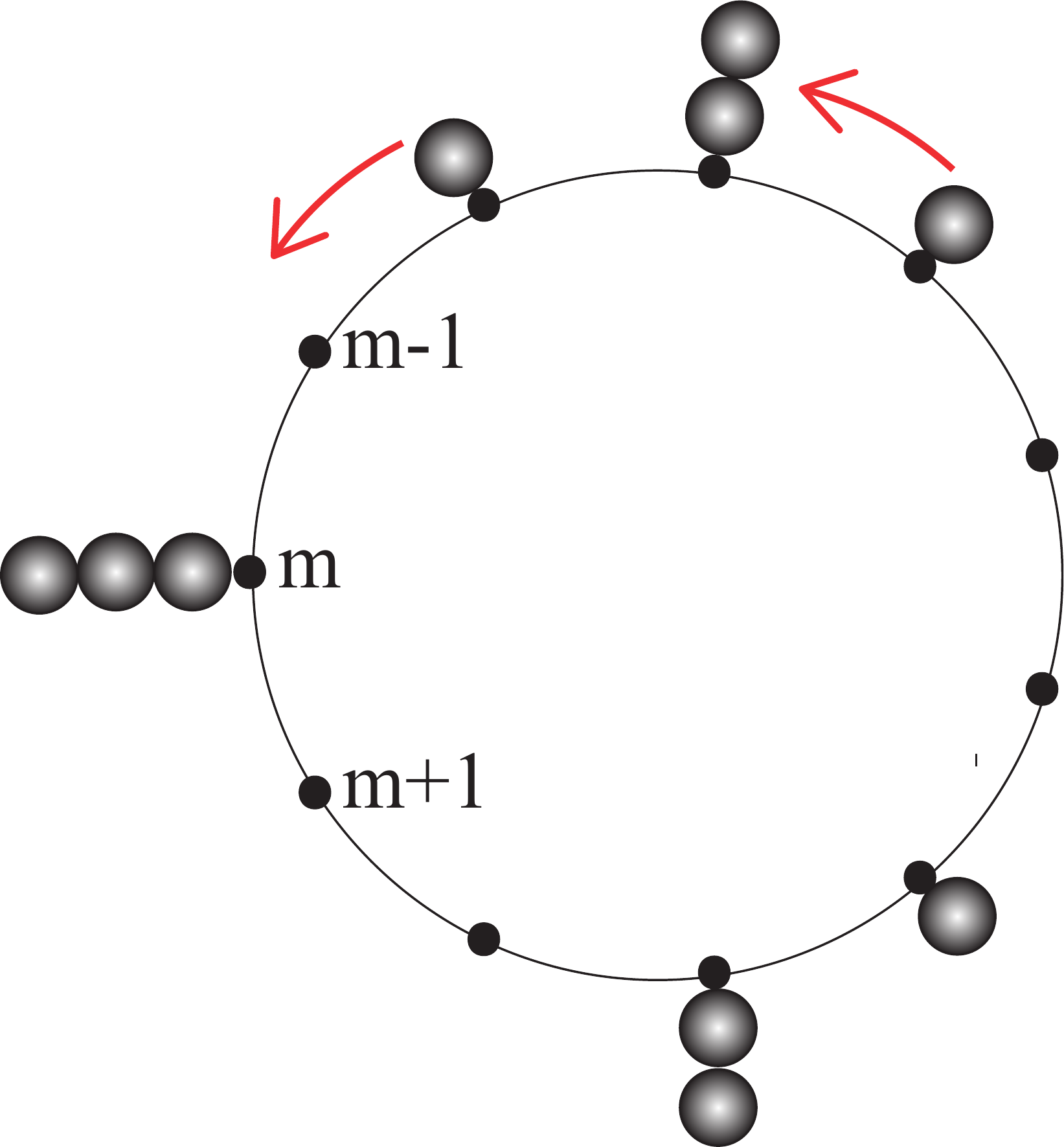}
\caption{Totally asymmetric simple zero range process.}\label{Fig1}
\end{figure}

A conf\/iguration $C$ of the system is characterized by the list of all possible arrangements of~$N$ particles amongst the $M+1$ available sites. The total number of conf\/igurations is, therefore,
\begin{gather}
\Omega =\frac{\left( N+M\right) !}{\left( N\right) !\left( M\right) !} .\label{nc}
\end{gather}

The probability $P_t(C)$ of f\/inding the system in conf\/iguration $C$ at time $t$ satisf\/ies the master equation
\begin{gather}
\frac{{\rm d}P_t(C)}{{\rm d}t}=\mathcal{M}P_t(C), \label{meq}
\end{gather}
where $\mathcal{M}$ is the Markov matrix of the size $\Omega \times \Omega $. For $C\neq C^{\prime }$ the entry $\mathcal{M}(C^{\prime },C)$ is the transition rate from $C$ to $C^{\prime }$. It is equal to unity, if the transition is allowed, and is zero otherwise. The diagonal entry $-\mathcal{M}(C,C)$ is equal to the number of \textit{occupied} sites in the conf\/iguration $C$. The elements of the columns and rows of $\mathcal{M}$ add up to zero and total probability is conserved
\begin{gather*}
\sum_CP_t(C)=1 . 
\end{gather*}
The process described is stochastic, the unique stationary state of this system being the one, in which all $\Omega $ dif\/ferent conf\/igurations $C$ have equal weight.

A conf\/iguration $C$ is represented by the sequence $(n_0,n_1,\ldots ,n_M)$ of occupation numbers $n_j$, satisfying the condition $0\leq n_0,n_1,\ldots, n_M\leq N$, $n_0+n_1+\dots +n_M=N$. We can rewrite the master equation~(\ref{meq}) for the TAZRP in the form
\begin{gather}
\frac{{\rm d}}{{\rm d}t}P_t(n_0,\ldots ,n_M) =\sum_{m=0}^MP_t(n_0,\ldots ,n_m+1,n_{m+1}-1,\ldots ,n_M) \nonumber \\
\hphantom{\frac{{\rm d}}{{\rm d}t}P_t(n_0,\ldots ,n_M) =}{} -KP_t(n_0,\ldots ,n_M), \label{meqe}
\end{gather}
where $K$ is the number of occupied sites in the conf\/iguration $(n_0,n_1,\ldots ,n_M)$, i.e., the number of $n_j\neq 0$. This equation has
to be supplemented by the condition that $P_t(n_0,\ldots ,n_M)=0$, if at least one of the $n_j<0$.

To formulate the dynamic of the system in terms of a quantum mechanical model we denote a particle conf\/iguration as a Fock vector $|n_0,\ldots,n_M\rangle $ and def\/ine a probability vector
\begin{gather*}
|P_t\rangle =\sum_{0\leq n_0,n_1,\ldots ,n_M\leq N \atop n_0+n_1+\cdots +n_M=N}P_t(n_0,\ldots ,n_M)|n_0,\ldots ,n_M\rangle, 
\end{gather*}
where $P_t(n_0,\ldots ,n_M)$ are the probabilities of conf\/iguration $(n_0,\ldots ,n_M)$. The generator of the master equation (\ref{meqe}) may be
written as
\begin{gather}\label{ham}
H=\sum_{m=0}^M \big( \phi_m \phi_{m+1}^{\dagger } - \phi_m^{\dagger } \phi_m \big) =\sum_{m=0}^M \big( \phi_{m+1}^{\dagger } - \phi_m^{\dagger}\big) \phi_m .
\end{gather}
Here, the phase operators $\phi_n, \phi_n^{\dagger}$ were introduced. They satisfy commutation relations
\begin{gather*}
\lbrack \hat N_i,\phi_j]=-\phi_i \delta _{ij} ,\qquad [\hat N_i, \phi_j^{\dagger }] = \phi_i^{\dagger } \delta_{ij} ,\qquad [\phi_i,\phi_j^{\dagger }] = \pi_i \delta_{ij} , 
\end{gather*}
where $\hat N_j$ is the number operator and $\pi _i=1-\phi _i^{\dagger }\phi_i$ is vacuum projector: $\phi_j \pi_j = \pi_j \phi_j^\dagger=0$. This algebra has a representation on the Fock space:
\begin{gather}
\phi_j | n_0,\ldots , 0_j,\ldots , n_M\rangle =0, \nonumber \\
\phi_j^{\dagger }| n_0, \ldots , n_j, \ldots , n_M \rangle =| n_0, \ldots,n_j+1, \ldots , n_M \rangle , \nonumber \\
\phi_j | n_0,\ldots , n_j, \ldots , n_M \rangle = | n_0, \ldots , n_j-1, \ldots, n_M \rangle , \label{fs}\\
\hat N_j| n_0, \ldots ,n_j , \ldots , n_M \rangle = n_j | n_0, \ldots , n_j, \ldots, n_M\rangle ,\nonumber \\
\pi_j | n_0, \ldots , 0_j, \ldots , n_M \rangle = | n_0, \ldots , 0_j, \ldots , n_M \rangle. \nonumber
\end{gather}
The states $|n_0, \ldots , n_M \rangle $ are orthogonal, $\langle p_0, \ldots, p_M | n_0, \ldots , n_M \rangle = \prod\limits_{l=0}^M \delta_{p_l n_l}$. The Fock vectors are generated from the vacuum state
\begin{gather}
|{\bf 0} \rangle \equiv\prod_{j=0}^M|0\rangle _j \label{vac}
\end{gather}
by action of rising operators $\phi_j^{\dagger }$
\begin{gather}\label{state}
|{\bf n} \rangle \equiv|n_0,\ldots , n_M \rangle =\prod_{j=0}^M \bigl( \phi_j^{\dagger }\bigr)^{n_j} | {\bf 0}\rangle , \qquad 0 \leq n\leq N, \qquad \sum_{j=0}^M n_j=N.
\end{gather}
The total number operator
\begin{gather*}
 \hat{N}=\sum_{j=0}^M N_j
\end{gather*}
commutes with the Hamiltonian (\ref{ham}):
\begin{gather}\label{numbcon}
 [H,\hat{N}]=0.
\end{gather}

The hopping term $\phi_{m+1}^\dag\phi_m$ of the Hamiltonian (\ref{ham}) annihilates the particle in the site $m$ and creates it in the site $m+1$:
\begin{gather*}
\sum_{m=0}^M \phi_m \phi_{m+1}^{\dagger } | n_0, \ldots, n_M\rangle = \sum_{m=0}^M|n_0, \ldots, n_{m}-1,n_{m+1}+1, \ldots , n_M \rangle .
\end{gather*}
The vacuum projector $\pi_m$ due to the def\/inition (\ref{fs}) acts on the Fock vector in the following way
\begin{gather}\label{vpv}
\pi_m |n_0,\ldots , n_M \rangle=|n_0, \ldots, n_M \rangle , \qquad \text{if} \quad n_m=0 ,
\end{gather}
and
\begin{gather*}
\pi_m |n_0,\ldots ,n_M \rangle=0 , \qquad \text{if} \quad n_m \neq0 .
\end{gather*}

The operator
\begin{gather*}
\hat K = \sum_{m=0}^M \phi_m^{\dagger }\phi_m= \sum_{m=0}^M( 1-\pi _m)
\end{gather*}
counts the number of occupied sites in the conf\/iguration $(n_1, n_2,\ldots , n_M)$
\begin{gather*}
\hat K|n_0,\ldots ,n_M\rangle =K|n_0,\ldots ,n_M\rangle .
\end{gather*}

The evolution of the quantum system $|P_t\rangle =e^{tH}|P_0\rangle $ is governed by the imaginary time Schr\"{o}dinger equation
\begin{gather*}
\frac {{\rm d}}{{\rm d}t}|P_t\rangle =H|P_t\rangle , 
\end{gather*}
which is equivalent to the master equation (\ref{meqe}) for the
probabilities $P_t(n_0,\ldots ,n_M)$ equal to the matrix elements
\begin{gather*}
P_t(n_0,\ldots ,n_M)=\langle n_0,\ldots ,n_M|P_t\rangle . 
\end{gather*}
The initial ($t=0$) probability distribution def\/ines the state $|P_0\rangle $
\begin{gather*}
|P_0\rangle =\sum_{0\leq n_0,n_1,\ldots ,n_M \leq N \atop n_0+n_1+\cdots +n_M=N}P_0(n_0,\ldots ,n_M)|n_0,\ldots ,n_M\rangle .
\end{gather*}
The state
\begin{gather*}
\langle S|=\sum_{0 \leq n_0, n_1, \ldots ,n_M\leq N \atop n_0+n_1+\cdots +n_M=N}\langle n_0,\ldots , n_M|=\langle 0|\left(
\sum_{j=0}^M \phi_j \pi_{j-1}\cdots \pi_1 \right)^N 
\end{gather*}
is a left eigenvector of the model, which obeys
\begin{gather}
\langle S|H=0. \label{sham}
\end{gather}
Correspondingly, $H$ has one right eigenvector with eigenvalue zero, which is associated with the state $|\Omega \rangle =\frac 1\Omega |S\rangle $: $H|\Omega \rangle =0$, where $\Omega $ is (\ref{nc}), $\langle S|\Omega \rangle =1$. The vector $\langle S|$ does not evolve in time and, therefore, corresponds up to normalization factor to a steady state distribution of the system
\begin{gather*}
\langle S|P_t\rangle =\langle S|P_0\rangle =1.
\end{gather*}

In this paper we shall calculate the conditional probability
\begin{gather}\label{aver}
 P_t({\bf n}|{\bf m})=\frac 1{\Omega }\langle n_0,\ldots ,n_M|e^{tH} |m_0,\ldots ,m_M\rangle ,
\end{gather}
which is equal to probability, that in a time $t$ the system will be in a pure state def\/ined by the occupation numbers $( n_0,\ldots ,n_M)$ provided that initially the system was prepared in a pure state $|P_0\rangle=|m_0,\ldots ,m_M\rangle$.

\section{Solution of the TASZRP}\label{section3}

To apply the scheme of the QIM to the solution of the Hamiltonian (\ref{ham}) we def\/ine $L$-ope\-ra\-tor~\cite{nas} which is $2\times2$ matrix with the operator-valued entries acting on the Fock states according to~(\ref{fs}):
\begin{gather}
L (n |u)\equiv \begin{pmatrix}
u^{-1}+u \pi_n & \phi_n^{\dagger } \\
\phi_n & u
\end{pmatrix} , \label{lopz}
\end{gather}
where $u\in \mathbb C$ is a parameter. This $L$-operator satisf\/ies the intertwining relation
\begin{gather*}
R(u,v)\left( L(n|u)\otimes L(n|v)\right) =\left( L(n|v)\otimes L(n|u)\right) R(u,v) , 
\end{gather*}
in which $R(u,v)$ is the $R$-matrix
\begin{gather}
R(u,v)=\begin{pmatrix}
f(v,u) & 0 & 0 & 0 \\
0 & g(v,u) & 1 & 0 \\
0 & 0 & g(v,u) & 0 \\
0 & 0 & 0 & f(v,u)
\end{pmatrix} , \label{r}
\end{gather}
where
\begin{gather}
f(v,u)=\frac{u^2}{u^2-v^2},\qquad g(v,u)=\frac{uv}{u^2-v^2} ,\qquad u, v\in \mathbb C .\label{fg}
\end{gather}

The monodromy matrix is the matrix product of $L$-operators
\begin{gather}
T(u)=L(M|u) L(M-1|u) \cdots L(0|u)=\begin{pmatrix}
A(u) & B(u) \\
C(u) & D(u)
 \end{pmatrix} . \label{mm}
\end{gather}
The commutation relations of the matrix elements of the monodromy matrix are given by the same $R$-matrix~(\ref{r})
\begin{gather}
R(u,v)\left( T(u)\otimes T(v)\right) =\left( T(v)\otimes T(u)\right) R(u,v) . \label{ttr}
\end{gather}
The transfer matrix $\tau (u)$ is the trace of the monodromy matrix in the auxiliary space
\begin{gather}
\tau (u)=\operatorname{tr} T(u)= A(u)+D(u).\label{trans}
\end{gather}
The relation (\ref{ttr}) means that $[\tau (u),\tau (v)]=0$ for arbitrary $u, v\in \mathbb C$.

From the def\/initions (\ref{lopz}) and (\ref{mm}) one f\/inds by direct calculation that the entries of the monodromy matrix are polynomials in~$u^2$. For $A(u)$ and $D(u)$ one has
\begin{gather}
u^{M+1} A(u) =  1+u^2 \left(\sum_{m=0}^{M-1} \phi_m \phi_{m+1}^{\dagger}+ \sum_{m=0}^{M} \pi_m \right) +\cdots + u^{2(M+1)} \prod_{m=0}^M \pi_m , \nonumber\\
u^{M+1} D(u) = u^2 \phi_0^{\dagger } \phi_M + \cdots +u^{2(M+1)} ,\label{reprad}
\end{gather}
where the dots stand for the terms not important for further consideration. We also f\/ind that
\begin{gather}
 \lim_{u\rightarrow 0} \widetilde B(u)\equiv \lim_{u\rightarrow 0} u^{M} B(u) =\phi_0^{\dagger } ,\label{reprbc1} \\
\lim_{u\rightarrow 0} \widetilde C(u)\equiv \lim_{u\rightarrow 0} u^{M}C(u) =\phi _M . \label{reprbc}
\end{gather}
The representation (\ref{reprad}) allows to express the Hamiltonian (\ref{ham}) through the transfer mat\-rix~(\ref{trans})
\begin{gather}
H=\frac{\partial }{\partial u^2} u^{M+1}\tau (u) \Bigl|_{u=0}-(M+1)=\frac{\partial }{\partial u^2} u^{M+1} ( A(u)+D(u))\Bigl|_{u=0}-(M+1) . \label{hamrep}
\end{gather}
By construction this Hamiltonian commutes with the transfer matrix
\begin{gather*}
\lbrack H,\tau (u)]=0.
\end{gather*}

Since the Hamiltonian (\ref{ham}) is non-Hermitian we have to distinguish between its right and left eigenvectors. The $N$-particle right state-vectors are taken in the form
\begin{gather}
|\Psi_N ({\bf u})\rangle =\left(\prod_{j=1}^N \widetilde{B}(u_j) \right)|{\bf 0} \rangle, \label{b}
\end{gather}
where $\widetilde{B}(u)$ is def\/ined in (\ref{reprbc1}), and ${\bf u}$ implies a collection of arbitrary complex parameters $u_j\in \mathbb C$:
${\bf u}=(u_0, u_1, \ldots, u_N)$. The left state-vectors are equal to
\begin{gather}
\langle \Psi_N ({\bf u})|=\langle {\bf 0} | \left(\prod_{j=1}^N \widetilde{C}(u_j)\right) , \label{c}
\end{gather}
where $\widetilde{C}(u)$ is given by (\ref{reprbc}). The vacuum state (\ref{vac}) is an eigenvector of~$A(u)$ and~$D(u)$,
\begin{gather*}
A(u)|{\bf 0} \rangle =\alpha (u)| {\bf 0}\rangle, \qquad D(u)|{\bf 0} \rangle =\delta (u)|{\bf 0} \rangle \label{advz}
\end{gather*}
with the eigen-values
\begin{gather}
\alpha (u)=\big(u^{-1}+u\big)^{M+1},\qquad \delta (u)=u^{M+1}. \label{evz}
\end{gather}

The state-vectors (\ref{b}) and (\ref{c}) are the eigenvectors both of the Hamiltonian~(\ref{ham}) and of the transfer matrix~$\tau(u)$~(\ref{trans}), if, and only if, the variables $u_j$ satisfy the Bethe equations
\begin{gather*}\label{beqgf}
\frac{\alpha (u_n)}{\delta (u_n)}=\prod_{m\neq n}^N \frac{f(u_m, u_n)}{f(u_n, u_m)} ,
\end{gather*}
where $f$ are the elements of the $R$-matrix~(\ref{fg}). In the explicit form the Bethe equations are given by
\begin{gather}
u_n^{-2N} \big(1+ u_n^{-2}\big)^{M+1} =\frac{(-1)^{N-1}}{U^2} ,\qquad U^2\equiv \prod_{j=1}^N u_j^{2} .
\label{bethez}
\end{gather}
There are $\Omega$ equation~(\ref{nc}) sets of solutions of these equations. The eigenvalues $\Theta_N (v, {\bf u})$ of the transfer matrix~(\ref{trans}) in the general form are equal to
\begin{gather*}\label{etrm}
 \Theta _N (v; {\bf u})=\alpha (v) \prod_{j=1}^N f (v, u_j) + \delta (v) \prod_{j=1}^N f(u_j, v) .
\end{gather*}
For the model under consideration
\begin{gather*}
v^{M+1} \Theta_N (v; {\bf u})=\big(1+ v^2\big)^{M+1}\prod_{m=1}^N \frac{u_m^2}{u_m^2-v^2} + v^{2(M+1)} \prod_{m=1}^N \frac{v^2}{v^2-u_m^2}\\
\hphantom{v^{M+1} \Theta_N (v; {\bf u})}{} =\bigl(\big(1+ v^2\big)^{M+1} + (-1)^{N} v^{2(M+N+1)} U^{-2} \bigr) \mathcal{H}\big(v^{2}; {\bf u}^{-2}\big).
\end{gather*}
Here, the generating function of complete symmetric functions $h_{l}\big({\bf u}^{-2}\big)\equiv h_{l} \big(u_1^{-2}, u_2^{-2}, \ldots$, $u_N^{-2}\big)$~\cite{macd} is introduced
\begin{gather*}
 \mathcal{H}\big(v^{2}; {\bf u}^{-2}\big) \equiv \prod_{m=1}^N \frac{1}{1- v^2/ u_m^2} = \sum_{l \ge 0} h_{l}\big({\bf u}^{-2}\big) v^{2 l} .\label{evtmzr3}
\end{gather*}
Equation (\ref{hamrep}) enables to obtain the spectrum of the Hamiltonian~(\ref{ham}). The $N$-particle eigenenergies
\begin{gather*}
 H|\Psi_N ({\bf u})\rangle=E_N |\Psi_N ({\bf u})\rangle
\end{gather*}
are equal to
\begin{gather}\label{eienz}
 E_N({\bf u})=\frac{\partial }{\partial v^2} v^{M+1} \Theta_N (v; {\bf u}) \Bigl|_{v=0}  =h_{1} \big({\bf u}^{-2}\big) =\sum_{k=1}^N u_k^{-2} .
\end{gather}
The steady state (\ref{sham}) corresponds to a special solution of Bethe equations (\ref{bethez}) when all $u_j=\infty$.

\section{The calculation of conditional probability}\label{section4}

For the models associated with the $R$-matrix (\ref{r}) the scalar product of the state-vectors (\ref{b}) and (\ref{c}) is given by the formula \cite{bogol}:
\begin{gather}\label{scprgv}
 \langle\Psi_N ({\bf v})|\Psi_N ({\bf u})\rangle = \frac{\det Q}{{{\mathcal V}}_{N}\big({\textbf v}^2\big){{\mathcal V}}_{N}\big({\textbf u}^{-2}\big)} \prod_{j=1}^N \left(\frac{v_j}{u_j} \right)^{M+N-1} ,
\end{gather}
where ${{\mathcal V}}_{N}({\textbf x})$ is the Vandermonde determinant,
\begin{gather}
{{\mathcal V}}_{N}({\textbf x}) \equiv {{\mathcal V}}_{N}(x_1, x_2, \ldots, x_N) = \prod_{1\le i< k\le N} (x_k-x_i) ,\label{wander}
\end{gather}
and the matrix $Q$ is characterized by the entries $Q_{j k}$, $1\le j, k\le N$,
\begin{gather*}
 Q_{jk}=\frac{\displaystyle \alpha(v_j)\delta(u_k)\left( \frac{u_k}{v_j}\right)^{N-1} - \alpha(u_k) \delta(v_j)\left( \frac{v_j}{u_k}\right)^{N-1}} {\displaystyle \frac{u_k}{v_j}- \frac{v_j}{u_k}} ,
\end{gather*}
with $\alpha(u)$ and $\delta(u)$ given by (\ref{evz}).

The norm of the state-vector ${\cal N}^2({\bf u}) \equiv \langle \Psi_N ({\bf u})|\Psi_N ({\bf u})\rangle$ is def\/ined by the scalar product (\ref{scprgv}) when the arguments ${\bf v}$ and ${\bf u}$ satisfy the Bethe equations~(\ref {bethez}). For the present case of the generalized phase model we substitute $v_k=u_k$, $\forall\, k$, respecting the Bethe equations~(\ref{bethez}) into the entries of the matrix $Q$. The resulting matrix is denoted as $\widetilde Q$, and its entries at $j\ne k$ are equal to
\begin{gather*}
 \widetilde{Q}_{jk} = \frac{ (-1)^N  (u_k u_j)^{N+M+1}}{ U^{2}} ,
\end{gather*}
where $U^2$ is given by~(\ref{bethez}). L'H\^ospital rule gives the diagonal entries of~$\widetilde{Q}$
\begin{gather*}
 \widetilde{Q}_{j j} = (N-1) \alpha(u_j) \delta(u_j) + \bigl( \alpha(u_j) \delta'(u_j)-\alpha'(u_j) \delta(u_j)\bigr)\frac{u_j}{2} \\
\hphantom{\widetilde{Q}_{j j}}{} = \bigl(1-N-\mathfrak{G}_j \bigr) \frac{ (-1)^N u_j^{2(N+M+1)}}{ U^{2}} ,
\end{gather*}
where
\begin{gather*}\label{summa}
\mathfrak{G}_j \equiv \frac{M+1}{ u_j^2+1} .
\end{gather*}
As a result, the squared norm ${\cal N}^2({\bf u})$ on the Bethe solution takes the form
\begin{gather}
{\cal N}^2({\bf u}) = \frac{\det \widetilde Q}{ {{\mathcal V}}_{N}\big({\textbf u}^2\big){{\mathcal V}}_{N}\big({\textbf u}^{-2}\big)} , \label{norm1} \\
 \det \widetilde Q = U^{2(M+1)} \left(1- \sum_{l=1}^N \frac{1}{N+\mathfrak{G}_l}\right) \prod_{j=1}^N (N+ \mathfrak{G}_j) . \label{norm2}
\end{gather}

The state-vectors belonging to the dif\/ferent sets of solutions of the Bethe equations~(\ref {bethez}) are orthogonal. The eigenvectors~(\ref{b}) and~(\ref{c}) provide the resolution of the identity operator
\begin{gather}\label{ident}
I=\sum_{\{{\bf u}\}}\frac{ |\Psi_N ({\bf u})\rangle \langle \Psi_N ({\bf u})| }{{\cal N}^2({\bf u})},
\end{gather}
where the summation $\sum_{\{{\bf u}\}}$ is over all independent solutions of the Bethe
equations (\ref {bethez}).

Inserting the resolution of the identity operator (\ref{ident}) into (\ref{aver}), one obtains the general answer for the conditional probability
\begin{gather}\label{gadcf}
P_t({\bf n}\,|\,{\bf m})=\frac 1{ \Omega } \sum_{\{{\bf u}\}}e^{t E_N(\bf {u})} \frac{ \langle \mathbf{n}|\Psi_N ({\bf u})\rangle \langle \Psi_N
({\bf u})|\mathbf{m}\rangle }{{\cal N}^2({\bf u})} .
\end{gather}

For the simplicity let us consider the initial state equal to $|N, 0, \ldots, 0\rangle$ and the f\/inal one respectively to $\langle 0, 0, \ldots, N|$. The conditional probability (\ref{aver}) of this process is specif\/ied as follows
\begin{gather}\label{gfno}
P_t \equiv \frac 1{\Omega }\big\langle 0, 0, \ldots, N \big| e^{tH} \big| N, 0, \ldots, 0 \big\rangle =\frac 1{\Omega }\big\langle {\bf 0} \big| (\phi_M)^N e^{tH} (\phi^\dag_0)^N \big| {\bf 0} \big\rangle ,
\end{gather}
where equation~(\ref{state}) has been used. Inserting the resolution of the identity operator~(\ref{ident}) into~(\ref{gfno}), we obtain
\begin{gather*}
 P_t=\frac 1{\Omega }\sum_{\{{\bf u}\}}\frac{ e^{t E_N({\bf u}) }}{{\cal N}^2({\bf u})}
 \big\langle {\bf 0} | (\phi_M)^N | \Psi_N ({\bf u})\rangle \langle \Psi_N ({\bf u})| (\phi^\dag_0)^N | {\bf 0}\big\rangle ,
\end{gather*}
where the summation is over all independent solutions of equations~(\ref{bethez}).

The decomposition (\ref{reprbc}) for $B(u)$ and $C(u)$ gives that
\begin{gather*}
 \big\langle {\bf 0} | (\phi_M)^N | \Psi_N ({\bf u})\big\rangle =\lim_{\mathbf{v}\rightarrow 0}\langle \Psi_N ({\bf v})|\Psi_N ({\bf u})\rangle = 1 ,  \\
 \big\langle \Psi_N({\bf u})| (\phi^\dag_0)^N | {\bf 0}\big\rangle =\lim_{\mathbf{v}\rightarrow 0}\langle \Psi_N ({\bf u})|\Psi_N ({\bf v})\rangle = 1 ,
\end{gather*}
and eventually the answer is
\begin{gather*}
 P_t=\frac 1{\Omega }\sum_{\{{\bf u}\}}\frac{e^{tE_N(\bf{u})}}{{\cal N}^2({\bf u})} ,
\end{gather*}
where ${{\cal N}^2({\bf u})}$ is given by (\ref{norm1}), (\ref{norm2}).

\begin{figure}[h]\centering
\includegraphics [scale=1.20]{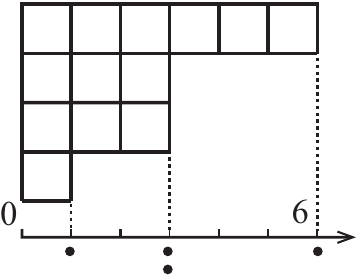}
\caption{A conf\/iguration of particles ($N = 4$) on a lattice ($M = 6$), the corresponding partition ${\boldsymbol{\lambda}} =(6^1, 5^0, 4^0, 3^2, 2^0, 1^1, 0^0)\equiv(6, 3, 3, 1)$ and its Young diagram.}\label{Fig2}
\end{figure}

To obtain the explicit answer for the conditional probability in the general case (\ref{gadcf}) we shall express state vectors (\ref{b}) and (\ref{c}) in the coordinate form. The state-vector (\ref{b}) has the representation
\begin{gather}\label{stveczrp}
|\Psi_N ({\bf u })\rangle=\sum_{{{{\boldsymbol{\lambda}} }}\subseteq \{M^N\}}\chi^R_{{\boldsymbol{\lambda}}}({\bf u}) \left(\prod_{j=0}^M( \phi_{j}^{\dag })^{n_j}\right) | {\bf 0}\rangle ,
\end{gather}
where the symmetric function $\chi^R_{{\boldsymbol{\lambda}}}$ is equal, up to a multiplicative pre-factor, to
\begin{gather}\label{gf}
\chi^R_{{\boldsymbol{\lambda}}}({\bf x})= \chi^R_{{\boldsymbol{\lambda}}}(x_1, x_2, \ldots, x_N)=\frac{1}{{\mathcal V}_N({\bf x})} \det
\left( \frac{x_i^{2(N-j)}}{(1 + x_i^{-2})^{\lambda_j}} \right)_{1\le i, j\le N} .
\end{gather}
Here ${\boldsymbol{\lambda}}$ denotes the partition $(\lambda_1, \ldots, \lambda_N)$ of non-increasing non-negative integers,
\begin{gather*}M\geq \lambda_1\geq \lambda_2 \geq \cdots \geq \lambda_N \geq 0,\end{gather*}
and ${\mathcal V}_N ({\bf x})$ is the Vandermonde determinant~(\ref{wander}). There is a one-to-one correspondence between a sequence of the occupation numbers $(n_0, n_1, \ldots , n_M)$, $n_0+n_1+\dots+n_M=N$, and the partition
\begin{gather*}{\boldsymbol{\lambda}}=\big(M^{n_M}, (M-1)^{n_{M-1}},\ldots , 1^{n_1}, 0^{n_0}\big),
\end{gather*}
where each number $S$ appears $n_S$ times (see Fig.~\ref{Fig2}). The sum in equation~(\ref{stveczrp}) is taken over all partitions ${{\boldsymbol{\lambda}}}$ into at most $N$ parts with $N\leq M$.

Acting by the Hamiltonian (\ref{ham}) on the state-vector (\ref{stveczrp}), we f\/ind that the wave function (\ref{gf}) satisf\/ies the equation
\begin{gather}\label{wfeq}
 \sum_{k=1}^N \chi^R_{\lambda_1, \ldots, \lambda_k+1, \ldots, \lambda_N}({\bf u})=E_N({\bf u}) \chi^R_{\lambda_1, \ldots, \lambda_N}({\bf u}) ,
\end{gather}
together with the exclusion condition
\begin{gather}\label{exc}
 \chi^R_{\lambda_1, \ldots, \lambda_{l-1}=\lambda_l-1, \lambda_l, \ldots, \lambda_N} ({\bf u})= \chi^R_{\lambda_1, \ldots, \lambda_{l-1}= \lambda_l, \lambda_l, \ldots, \lambda_N}({\bf u}) ,  \qquad 1\le l\le N.
\end{gather}
The energy $E_N$ is given by (\ref{eienz}). The state-vector~(\ref{stveczrp}) is the eigenvector of the Hamilto\-nian~(\ref{ham}) with the periodic boundary conditions if the parameters $u_j$ satisfy the Bethe equations~(\ref{bethez}).

The relations (\ref{stveczrp}), (\ref{wfeq}) and (\ref{exc}) can be viewed as an implementation of the coordinate Bethe ansatz~\cite{kbi}, which is an alternative to the approach of the algebraic Bethe ansatz considered in Section~3. Although the model is solved by the algebraic Bethe ansatz, representations of the type of~(\ref{stveczrp}) are especially useful in discussing the combinatorial
properties of the quantum integrable models~\cite{b, xxz3, bmumn}.

Expanding the left state-vector (\ref{c}), we obtain
\begin{gather}\label{stveczrpl}
 \langle\Psi_N ({\bf u})|=\sum_{{{{\boldsymbol{\lambda}}}}\subseteq \{M^N\}}\chi^L_{{\boldsymbol{\lambda}}}({\bf u}) \langle {\bf 0} \big| \left(\prod_{i=0}^M\phi_i^{n_i}\right) ,
\end{gather}
where the wave function is given by the symmetric function
\begin{gather*} 
 \chi^L_{{\boldsymbol{\lambda}}}({\bf x})= \frac{\det\big(\big( 1+x_i^{-2} \big)^{\lambda_j}
 x_i^{2(N-j)} \big)_{1\le i, j\le N}}{{\mathcal V}_N({\bf x})} .
\end{gather*}
It satisf\/ies the equations
\begin{gather*}
\sum_{k=1}^N \chi^L_{\lambda_1, \ldots, \lambda_k-1, \ldots, \lambda_N}({\bf u})=E_N({\bf u})\chi^L_{\lambda_1, \ldots, \lambda_N}({\bf u}) ,\\
\chi^L_{\lambda_1, \ldots, \lambda_l, \lambda_{l+1}= \lambda_l+1, \ldots, \lambda_N} ({\bf u})= \chi^L_{\lambda_1, \ldots, \lambda_l, \lambda_{l+1}= \lambda_l, \ldots, \lambda_N} ({\bf u}) ,\qquad 1\le l\le N.
\end{gather*}

From equations~(\ref{stveczrp}), (\ref{stveczrpl}) one obtains
\begin{gather*}
\langle n_0, n_1, \ldots, n_M | \Psi_N ({\bf u}) \rangle= \chi^R_{{\boldsymbol{\lambda}}_R} ({\bf u}), \\
\langle \Psi_N ({\bf u})| m_0, m_1, \ldots, m_M \rangle=\chi^L_{{\boldsymbol{\lambda}}_L} ({\bf u}) ,
\end{gather*}
where
\begin{gather*}{\boldsymbol{\lambda}}_R= \big(M^{n_M}, (M-1)^{n_{M-1}}, \ldots ,1^{n_1}, 0^{n_0} \big) ,\qquad {\boldsymbol{\lambda}}_L=\big(M^{m_M}, (M-1)^{m_{M-1}}, \ldots , 1^{m_1}, 0^{m_0} \big) .\end{gather*}
Finally, the expression for the conditional probability (\ref{gadcf}) has the form
\begin{gather}\label{zrgf}
 P_t(\mathbf{n}\,|\,\mathbf{m})=\frac{1}{\Omega} \sum_{\{{\bf u}\}} \frac{ e^{t E_N ({\bf u})}}{{\cal N}^2 ({\bf u})} \chi^R_{{\boldsymbol{\lambda}}_R} ({\bf u}) \chi^L_{{\boldsymbol{\lambda}}_L} ({\bf u}) .
\end{gather}
Here ${\cal N}^2({\bf u})$ is the squared norm (\ref{norm1}).

\section{Multi-dimensional lattice walks bounded by a hyperplane}\label{section5}

\begin{figure}[h]\centering
\includegraphics[scale=.80]{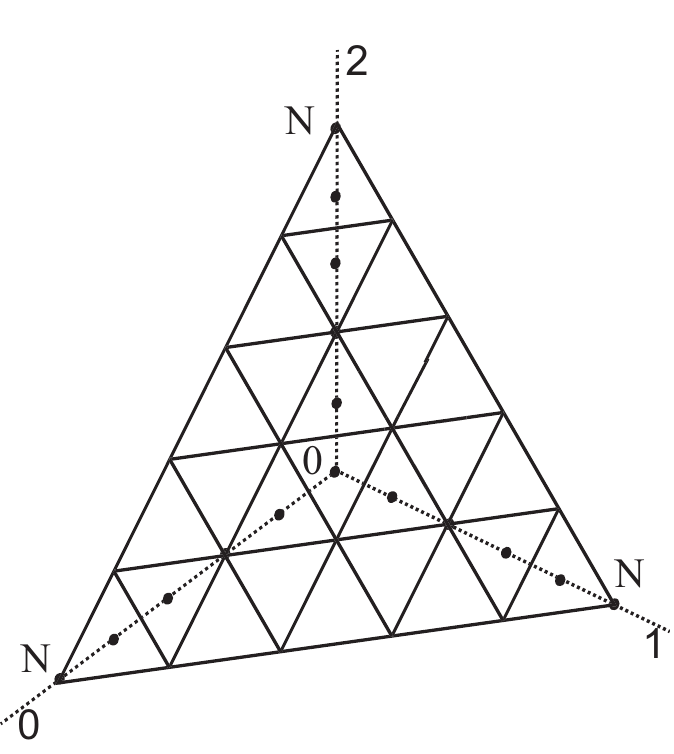}
\caption{A two-dimensional triangular simplicial lattice.}\label{Fig3}
\end{figure}

Starting from $(M+1)$-dimensional hypercubical lattice with unit spacing $\mathbb{Z}^{M+1}\ni \mathbf{m}\equiv (m_0, m_1,$ $\ldots, m_M)$, let us def\/ine the non-negative orthant $\mathbb{N}_0^{M+1}\equiv \{\mathbf{m} \,|\, 0\le m_{i}, i\in {{\mathcal M}}\}$ as a subset of $\mathbb{Z}^{M+1}$ (hereafter ${{\mathcal M}}\equiv \{0, 1, \ldots M\}$). Consider a subset of $\mathbb{N}_0^{M+1}$ consisting of sites with coordinates constrained by the requirement $m_0+m_1+\cdots+m_M = N$:
\begin{gather*}
{\rm Simp}_{(N)}(\mathbb{Z}^{M+1}) \equiv \left\{ {\bf m} \in \mathbb{N}_0^{M+1} \Big | \sum_{i\in {{\mathcal M}}} m_i = N \right\}. 
\end{gather*}
The set ${\rm Simp}_{(N)}\big(\mathbb{Z}^{M+1}\big)$ is compact $M$-dimensional, and we shall call it \textit{simplicial lattice}. A~two-dimensional triangular simplicial lattice is presented in Fig.~\ref{Fig3}. A~sequence of~$K+1$ points in~$\mathbb{Z}^{M+1}$ is called \textit{lattice path} of $K$ steps~\cite{kratt}.

Random walks over sites of ${\rm Simp}_{(N)}\big(\mathbb{Z}^{M+1}\big)$ are def\/ined by a set of admissible steps $\Omega_M$ (\textit{step set} $\Omega_M$) so that at each step an $i^{{\rm th}}$ coordinate $m_i$ increases by unity, while the nearest neighbouring one decreases by unity. Namely, each element of $\Omega_M$ is given by sequence $(e_0, e_2, \ldots, e_M)$ so that $e_i=\pm1$, $e_{i+1}=\mp1$ for all pairs $(i, i+1)$ with $i\in {{\mathcal M}}$ and $M+1=0$ $({\rm mod}\, 2)$, and $e_j=0$ for all $j\in {{\mathcal M}}$ and $j\neq i, i+1$. The step set $\Omega_M\equiv \Omega_M({\bf m}_0)$ ensures that trajectory of a random walk (lattice path) determined by the starting point $\mathbf{m}_0$ lies in $M$-dimensional set ${\rm Simp}_{(N)}\big(\mathbb{Z}^{M+1}\big)$.

\begin{figure}[h]\centering
\includegraphics [scale=.5] {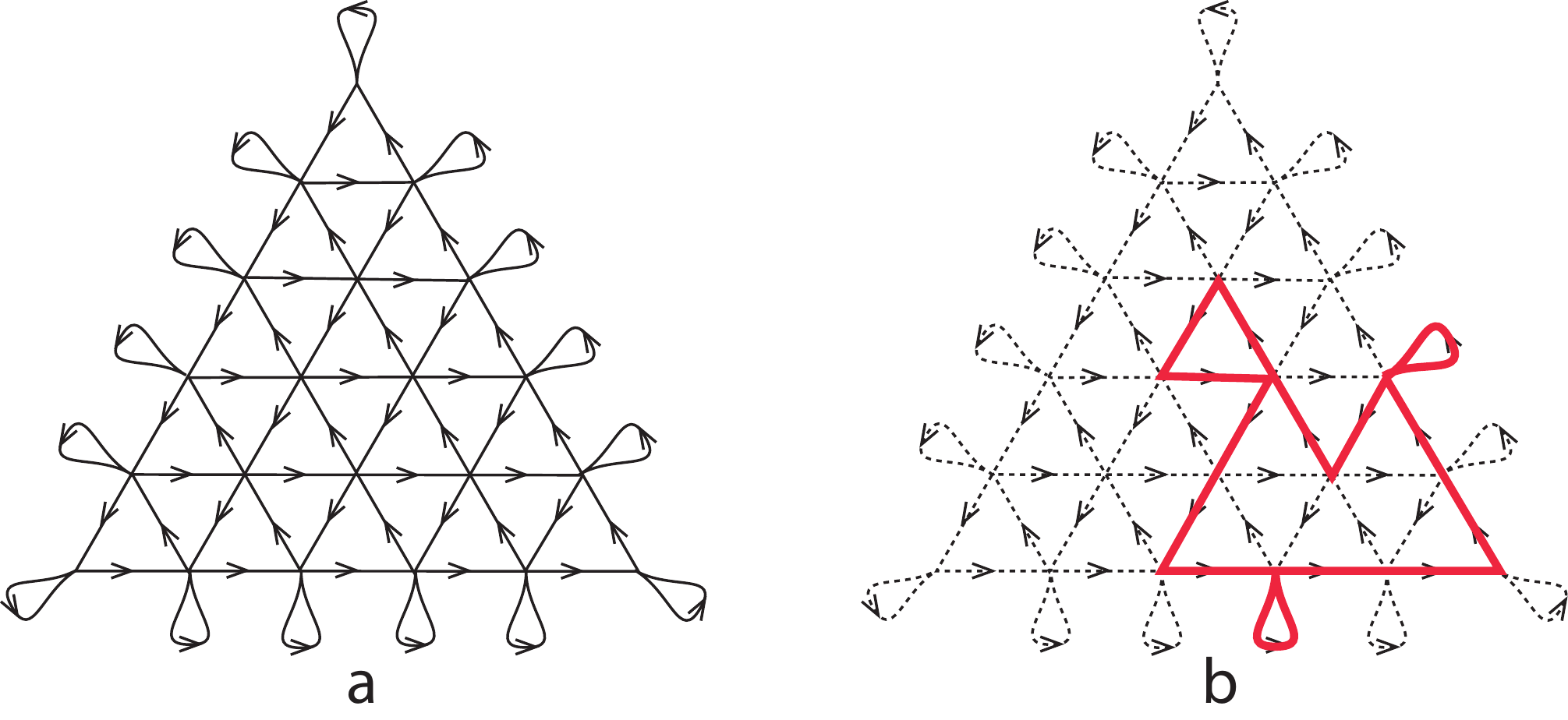}
\caption{The oriented two-dimensional simplicial lattice def\/ined by the step set $\Gamma_2$ with retaining boundary conditions~(a). The round trip lattice path of the $14$ steps from the node with coordinates $(0,5,0)$ is shown in~(b).}\label{Fig4}
\end{figure}

\textit{Directed random walks} on $M$-dimensional \textit{oriented simplicial lattice} are def\/ined by a step set $\Gamma_M=(k_0, k_1,\ldots,k_M)$ so that $k_i=-1$, $k_{i+1}=1$ for all pairs $(i,i+1)$ with $i \in {{\mathcal M}}$ and $M+1=0$ $({\rm mod}\, 2)$, and $k_j=0$ for all $j\in{{\mathcal M}}\backslash \{i, i+1\}$. It may occur that some points on the boundary of the simplicial lattice also belong to a random walk trajectory. Therefore, the walker's movements should be supplied with appropriate boundary conditions. The boundary of the simplicial lattice consists of $M+1$ faces of highest dimensionality $M-1$. Under the \textit{retaining boundary conditions} the walker comes to a node of the boundary, and either continues to move in accordance with the elements of $\Gamma_M$, or keeps staying in the node. An oriented two-dimensional simplicial lattice with the retaining boundary conditions is presented in Fig.~\ref{Fig4}.

To establish the connection of the exponential generation function of lattice paths and the conditional probability~(\ref{aver}) we shall interpret the coordinates $n_j$ of a walker $\mathbf{n}=(n_0, n_1, \ldots, n_M)$ $\in \mathbb{Z}^{M+1}$ in a simplicial lattice ${\rm Simp}_{(N)}\big(\mathbb{Z}^{M+1}\big)$ as the occupation numbers of $(M+1)$-component Fock space and describe the steps of a walker with the help of the Fock state-vectors $| {\bf n} \rangle \equiv$ $| n_0, n_1, \ldots, n_M \rangle$. Operator $\phi_j$ shifts the value of the $j^{\rm th}$ coordinate of the walker downwards $n_j\rightarrow n_j-1$, while $\phi_j ^{\dagger}$ upwards $n_j\rightarrow n_j+1$. If the walker arrived at an arbitrary node of $s^{\rm th}$ component of the boundary, it can make either allowed step or stay on it due the action of opera\-tor~$\pi_s$~(\ref{vpv}). The number operator $N_j$ acts as the coordinate operator $N_j|n_0,n_1,\ldots,n_j,\ldots,n_D\rangle =n_j|n_0,n_1$, $\ldots,n_j,\ldots,n_D\rangle$. Since the occupation numbers are non-negative integers and their sum is conserved (\ref{numbcon}), we can regard operator
\begin{gather}\label{rpgen}
H_{{\rm rw}}=\sum_{m=0}^M \bigl( \phi_m \phi_{m+1} ^{\dagger } + \pi_m\bigr)
\end{gather}
 as a generator of steps of a walker in ${\rm Simp}_{(N)}\big(\mathbb{Z}^{M+1}\big)$ with the retaining boundary conditions. The operators~(\ref{rpgen}) and~(\ref{ham}) are related
 \begin{gather*}
 H_{{\rm rw}}=H+M+1 .
 \end{gather*}

The number of lattice paths made by a walker over the $M$-dimensional simplicial lattice in $k$ steps with the ending nodes $(j_0,j_1,\ldots,j_M)$ and $ (l_0,l_1,\ldots,l_M)$ is given by the expression
\begin{gather}\label{numbpath}
 G_k({\bf l}\,|\,{\bf j})=\big\langle l_0,l_1,\ldots,l_M\big|H_{{\rm rw}}^k\big|j_0,j_1,\ldots,j_M \big\rangle .
\end{gather}
It is straightforward to verify that
\begin{gather*}
 G_k(\mathbf{l}\,|\,\mathbf{j}) = \sum_{s=0}^M G_{k-1}(\mathbf{l}\,|\, j_0, j_1, \ldots, j_s+1,j_{s+1}-1, \ldots, j_M) \\
  \hphantom{G_k(\mathbf{l}|\mathbf{j}) =}{} + (M+1-K) G_{k-1}(\mathbf{l}\,|\, j_0, \ldots, j_M) ,
\end{gather*}
where $k\ge 1$, and it is natural to impose the condition $G_0(\mathbf{l}\,|\,\mathbf{j})= \prod\limits_{l=0}^M \delta_{l_l j_l}$.

The exponential generating function of lattice paths in ${\rm Simp}_{(N)}(\mathbb{Z}^{M+1})$ is def\/ined as a formal series
\begin{gather}\label{genf}
 F_t(\mathbf{l}\,|\, \mathbf{j})= \sum_{k=0}^\infty\frac{t^k}{k!}G_k (\mathbf{l}\,|\,\mathbf{j}) .
\end{gather}
Due to def\/inition (\ref{numbpath}) it may be expressed as
\begin{gather*}
 F_t(\mathbf{l}\,|\, \mathbf{j})=\left\langle \mathbf{l}\big|\sum_{k=0}^\infty\frac{t^k}{k!}H_{{\rm rw}}^k\big|\mathbf{j}\right\rangle  =\big\langle \mathbf{l}\big|e^{tH_{{\rm rw}}}\big|\mathbf{j}\big\rangle .
\end{gather*}
Taking into account the connection (\ref{numbpath}) we obtain the desired relation of the conditional probability (\ref{aver}) and generation function of lattice paths~(\ref{genf}):
\begin{gather*}
 P_t(\mathbf{l}\,|\, \mathbf{j})=\frac{e^{-t(M+1)}}{\Omega} F_t(\mathbf{l}\,|\, \mathbf{j}) .
\end{gather*}

From the expression (\ref{zrgf}) it follows that the number of lattice paths made by a walker in the $M$-dimensional simplicial lattice in $k$ steps with the ending nodes $(j_0,j_1,\ldots,j_M)$ and $ (l_0,l_1,\ldots,l_M)$ is equal to
\begin{gather*}
 G_k(\mathbf{l}\,|\,\mathbf{j})= \sum_{\{{\bf u}\}} \frac{ h_1^k\big({\bf u}^{-2}\big)}{{\cal N}^2 ({\bf u})} \chi^R_{{\boldsymbol{\lambda}}_R} ({\bf u}) \chi^L_{{\boldsymbol{\lambda}}_L} ({\bf u}) ,
\end{gather*}
where the function $h_1\big({\bf u}^{-2}\big)$ is introduced in~(\ref{eienz}).

\subsection*{Acknowledgements}
This work was supported by RFBR grant 16-01-00296. N.M.B.~acknowledges the Simons Center for Geometry and Physics, Stony
Brook University at which some of the research for this paper was performed.

\pdfbookmark[1]{References}{ref}
\LastPageEnding

\end{document}